\documentclass[useAMS,usenatbib]{mn2e}
\usepackage{graphicx}
\title[On the nature of  CP Pup]{On the nature of CP Pup}
\author[E. Mason et al.]{E. Mason$^{1,6}$\thanks{E-mail:
emason@oats.inaf.it} M. Orio$^{2,8}$, K. Mukai$^{3}$, A. Bianchini$^{4}$, D. de Martino$^{5}$, F. di Mille$^{4,9}$, R.E. Williams$^{6}$, T. Abbot$^{7}$ and R. de Propris$^{7}$\\
$^{1}$INAF-Osservatorio Astronomico di Trieste, via G.B. Tiepolo 11, 14343, Trieste, Italy \\
$^{2}$INAF-Osservatorio Astronomico di Padova, vicolo dell'Osservatorio 5, I-35122 Padova, Italy\\
$^{3}$NASA Goddard Space Flight Center, Greenbelt, MD 20771, USA\\
$^{4}$Dipartimento di Astronomia, University of Padova, vicolo dell'Osservatorio 3, I-35122 Padova, Italy\\
$^{5}$INAF-Osservatorio Astronomico di Capodimonte, salita Moiariello 16, 80131, Napoli, Italy\\
$^{6}$Space Telescope Science Institute, 3700 San Martin Drive, Baltimore, MD  21218,   USA\\
$^{7}$AURA/CTIO, Casilla 603, La Serena, Chile\\
$^{8}$ Department of Astronomy, University of Wisconsin, 475 N Charter St., Madison, WI 53704, USA\\
$^{9}$Australian Astronomical Observatory-Carnegie Observatories, Colina El Pino, Casilla 601 La Serena, Chile
}
\begin{document}

\date{Accepted ; Received ; in original form }

\pagerange{\pageref{firstpage}--\pageref{lastpage}} \pubyear{2002}

\maketitle

\label{firstpage}

\begin{abstract}
We present new X-ray and optical spectra of the old nova CP\,Pup (nova Pup 1942) obtained with Chandra and the CTIO 4m telescope. The X-ray spectrum reveals a multi-temperature optically thin plasma reaching a maximum temperature of 36$^{+19}_{-16}$\,keV absorbed by local complex neutral material. The time resolved optical  spectroscopy confirms the presence of the $\sim$1.47\,hr period, with cycle-to-cycle amplitude changes, as well as of an additional long term modulation which is suggestive  either of a longer period or of non-Keplerian velocities in the emission line regions. These new observational facts add further support to CP\,Pup as a magnetic cataclysmic variable (mCV). We compare the mCV and the non-mCV scenarios and while we cannot conclude whether CP\,Pup is a long period system, all observational evidences point at an intermediate polar (IP) type CV. 
\end{abstract}


\section{Introduction}

CP\,Pup exploded  in  1942 as  one  of the  brightest  and fastest (t$_3$=8\,days, Schaefer \& Collazzi 2010)  of the Galactic classical  novae.  The luminosity of the post-nova never returned to  the pre-outburst quiescent  value. The nova remained  in a sort of standstill and, according to Schaefer \& Collazzi (2010), CP\,Pup might still  be more than  five magnitudes  brighter than  before the outburst.  
Thanks to its relatively bright ``quiescent'' magnitude and outburst characteristics, CP\,Pup has been widely studied as a rather ``extreme'' example of a nova, one that really helps constrain the parameters used in the classical nova (CN) models. The main findings were: 1) an unstable spectroscopic period of $\sim$1.47\,hr (Bianchini et al. 1985 and 2012, Duerbeck et al. 1987, O'Donoghue et al. 1989, White et al. 1993); 2) a $\sim$2 to 11\% longer (and unstable) photometric period (Warner 1985, Diaz \& Steiner 1991, Patterson \& Warner 1998); 3) an X-ray spectrum similar to  those of magnetic CV (mCV, Orio et al. 2009), 4) a modulation of the X-ray flux at the spectroscopic period (Orio et al. 2009, Balman et al. 1995); 5) double humped NIR light curve with period roughly matching the photometric period (Szkody \& Feinswog 1988).  
Taken together, these findings do not deliver a consistent picture of the old-nova (see details in Sect.5), and, in particular, leave unresolved the debate upon the nature of the nova (whether magnetic or non-magnetic) and its mass (too small to be consistent with either classical nova and/or stellar evolution theories). 

In this paper we present new {\sl Chandra} X-ray data and time resolved optical spectroscopy which we collected with the aim of addressing those inconsistencies. 

\section{Observations and data reduction} 

\subsection{The Chandra observations}
In the previous X-ray observation with {\sl XMM-Newton} (Orio et al. 2009), the upper limit of the RGS grating (0.33--2.4\,keV) 
 left an uncertainty concerning the highest plasma temperature in
 the cooling flow model with which we fitted the hot plasma. This temperature indicates the depth of the gravitational potential well of the white dwarf (WD), and, therefore, a way to measure the WD mass. In order to better constrain the plasma temperature and the WD mass  
 we observed CP\,Pup with {\sl Chandra} ACIS-S camera coupled with the High Energy Transmission Grating Spectrometer
(HETGS). The HETGS is a system of two sets of gratings, the HEG (high energy grating) and the MEG (medium energy grating). This allows  high resolution spectra (with E/$\Delta$E up to 1000) between 0.4\,keV and 10.0\,keV. CP\,Pup was observed between September 30 and October 31 2009: the total observing time was 175,780\,s, split in 7 exposures because of scheduling constraints. Table\,1 reports the log of observations together with the average count rate recorded during each exposure. 

\begin{table}
\scriptsize
\centering
\caption{\label{tab1} The seven {\sl Candra} exposures and their average count rate. Start and end times are in Terrestrial Time (TT).} 
\begin{tabular}{lcccccc} 
\hline 
Obs.ID &   Date      & Start & End   & Exposure  &  HEG &  MEG \\
      &             &       &       & (s)       &  counts/s & counts/s \\
\hline
9967  &  2009-09-30 & 15:25 & 21:29 & 19814 & 0.0196 & 0.0356 \\
11990 &  2009-10-02 & 05:23 & 19:20 & 47263 & 0.0207 & 0.0360 \\
9966  &  2009-10-11 & 08:30 & 16:01 & 24729 & 0.0198 & 0.0367 \\
12005 &  2009-10-15 & 09:53 & 19:41 & 32504 & 0.0237 & 0.0430 \\
12012 &  2009-10-26 & 00:43 & 06:53 & 19715 & 0.0246 & 0.0482 \\
11991 &  2009-10-28 & 14:44 & 20:52 & 19714 & 0.0252 & 0.0449 \\
12013 &  2009-10-31 & 06:28 & 09:45 & 9187  & 0.0225 & 0.0445 \\
\hline
\end{tabular}
\end{table}

\subsection{The CTIO observations}
Spectroscopic observations were taken at the CTIO's 4m telescope equipped with the, now retired, R-C spectrograph. The instrument setup was grating KPGL1-1 at a tilt angle of $\sim$57$^\circ$ and slit width of $\sim$1.2\,arcsec which, together, covered the wavelength range 3500--6000\,\AA \ at a resolution of $\sim$6\,\AA\footnote{This is the measured FWHM of the sky emission line.} (i.e. $\sim$320\,km/s). 
The exposures time was 60\,s and the duty cycle was $\sim$1.8\,min. A few times during each night the observing sequence was interrupted in order to take a wavelength calibration exposure. 
CP\,Pup was observed on Feb 6 and 7 2009 for about 5.9 and 4.7 consecutive hours, respectively. 
In total we collected 177 spectra during the first night and 125 during the second. The wavelength calibration exposures were used to calculate the dispersion solution of the spectra, however the wavelength ``zero-point'' was determined using the [O\,{\sc i}] sky line at 5577\,\AA, due to the spectrograph instability.  Each spectrum was corrected also for the earth motion, in addition to the instrumental flexure. 

The data were reduced following the standard procedure and using the dedicated IRAF packages. All the spectra were flux calibrated using the spectrophotometric standard stars HR4468 \& LTT4861 and LTT2415 for the first and the second night of observation, respectively. 

\section{Analysis of the Chandra data}

\begin{figure*}
\centering 
\includegraphics[width=14.truecm, angle=270]{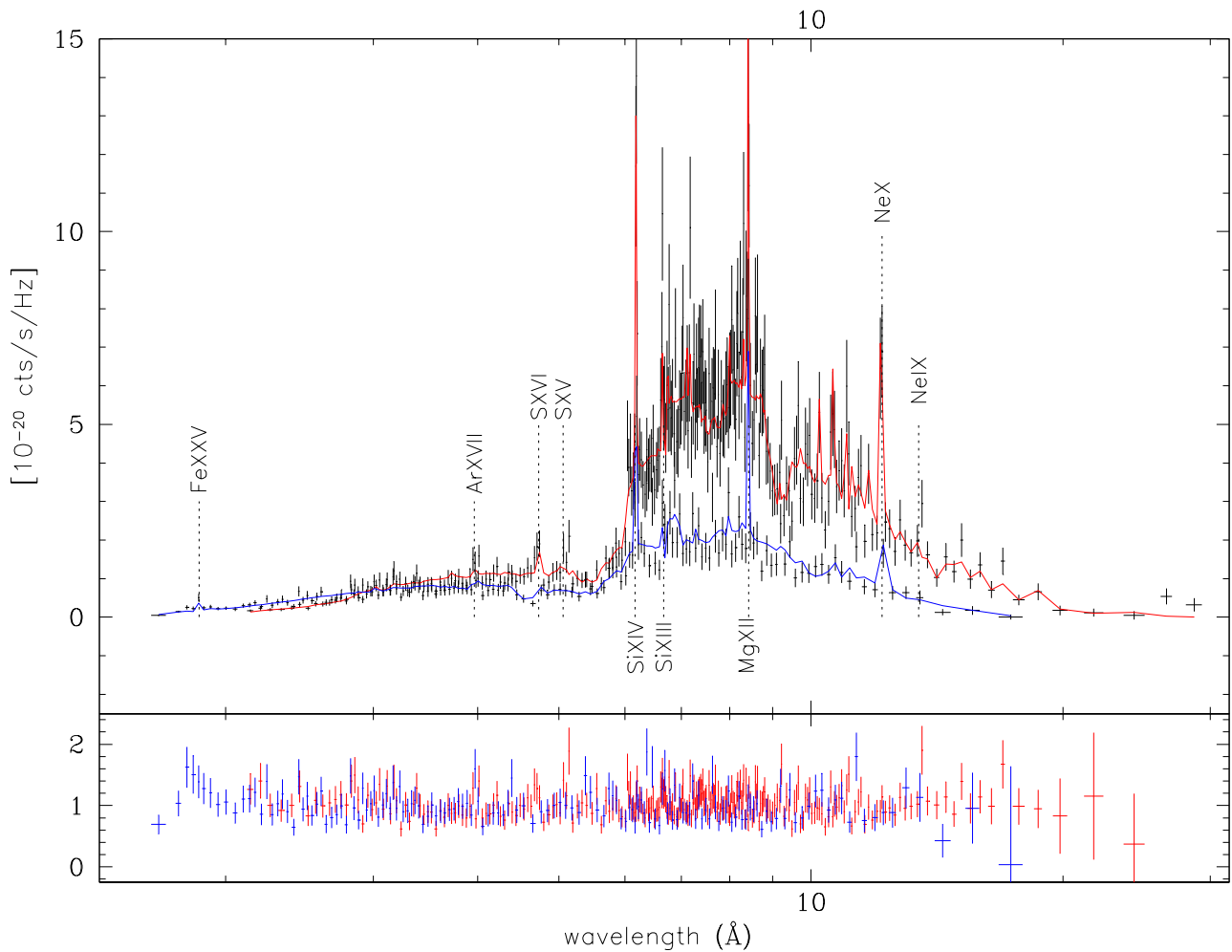}
\caption{Top panel: MEG data and their best fit model (red line) \& HEG data and their best fit model (blue line). Bottom panel: data/model ratio, with the same color convention as above. Note that the wavelength scale is logarithmic.}
\label{chandra}  
\end{figure*} 

We reprocessed all the HETG data using {\tt chandrarepro} in the CIAO v4.3 software package. Using the resulting pha2 files, with standard source and background regions, we extracted type\,I standard pha files by means of the {\tt ftool cmppha} for the source and the background, and combined the $\pm1^{st}$ orders into a single MEG file and a single HEG file, each with associated response and background (making use of the tool {\tt addascaspec}). We then combined these files from the 7 pointings into a single HEG spectrum and a single MEG spectrum.  We focused on the ranges 0.4--6 keV (2.07--31.00\,\AA) for the MEG and 0.7--8 keV (1.55-17.71\,\AA) for the HEG.
As the summed spectra have a modest statistical quality, we produced two different versions of the spectra with different binning, using the {\tt ftool grppha}. In particular, one was heavily re-binned so that each channel had at least 25 source photons; we fitted these files using $\chi^2$ statistic, which is well suited to evaluating the goodness of fit. We also used this version to produce the final plots in Fig.\ref{chandra}.  For the final fitting, done with XSPEC, v12.6, we used a binning by a factor of 2, and the C statistic (Cash, 1979), with the background-subtracted spectra.

The spectra clearly contain K shell emission lines of various species from Fe to O, as expected for X-ray emission from a multi-temperature plasma (Fig.~\ref{chandra}).  The strongest lines are those of He-like Fe at 1.86\,\AA, He-like Ar at 3.96\,\AA, H and He-like S at 4.73\,\AA, and 5.06\,\AA, respectively, H and He-like Si at 6.17\,\AA, and 6.67\,\AA, respectively, H like Mg at 8.43\,\AA, and H and He like Ne at 12.16\,\AA, and 13.47\,\AA, respectively. Hence, following Mukai et al. (2003), we modeled the spectra using a cooling flow model, and fixed the minimum temperature to the lowest value available in the model, kT=0.0808 keV. This is effectively zero, as cooler plasma does not contribute with significant emission in the HETG band.

In fitting the models with XSPEC, we started by using a {\tt phabs} ({\tt mkcflow+ga}) model, with the Gaussian component representing the fluorescent Fe line at 6.4\,keV (1.86\,\AA).  However, the fit was poor so that we applied a series of modifications to the model. We first added a partial covering absorber improving the fit significantly. Second, since the line strengths deviated from the predictions of a solar abundance cooling flow model, we used the variable abundance version of the cooling flow model ({\tt vmcflow}) and allowed the abundances of O, Ne, Mg, Si, S, and Fe to vary (we tied the Ni abundances to Fe). Finally, we applied the {\tt gsmooth} convolution model to the {\tt vmcflow} component with its second parameter set to 1.0 (constant velocity width) to broaden the model line widths to match the observations. 

Applying the best-fit model from the C statistic fit to the heavily grouped spectra, we obtained a reduced $\chi^2$ of 1.1048 for 416 degrees of freedom (and 1.0870 for 413 degrees of freedom above 0.5 keV). Many of the fit parameters (27 in total\footnote{They include the elemental abundances for 14 atoms, minimum and maximum temperature, column density of the simple and partial absorber, redshift, normalization/\.{m}, etc.}) are poorly constrained. In particular, we cannot derive detailed elemental abundance but rather get the abundance pattern, which is suggestive of Ne abundance about twice solar (i.e. 2.2$\pm$0.6). We also derive the column density for the simple absorber and the partial covering absorbers:  N$_H$=0.16($\pm$0.03)$\times10^{22}$\,cm$^{-2}$ and 3.0($\pm$0.66)$\times 10^{22}$\,cm$^{-2}$, respectively; with a covering fraction of 45\% for the partial covering absorber.  We notice that the H column density measured by LAB and GLASS from the 21\,cm transition (Kalberla et al. 2005, Kalberla et al. 2010) in the direction of CP\,Pup is 0.672-0.673$\times10^{22}$\,cm$^{-2}$, i.e. significantly smaller than our total column density estimated by modeling of the Chandra spectra. This is an independent demonstration that we are witnessing local absorption in CP\,Pup, consistently with a complex local absorber. The model (absorbed) flux is 4.0$\times10^{-12}$\,erg/cm$^2$/s in the range 2--10\,keV and 4.8$\times10^{-12}$\,erg/cm$^2$/s in the range 0.4--10\,keV. This corresponds to 4.5$\times10^{-12}$\,erg/cm$^2$/s and 6.9$\times10^{-12}$\,erg/cm$^2$/s, respectively, once corrected for absorption, and to a luminosity of 1.4e33\,erg/s (2-10\,keV) and 2.1e33\,erg/s (0.4-10\,keV) for an assumed distance of 1.6 kpc (Williams 1982). For the same distance, the normalization constant in the cooling flow model indicates an accretion rate of 4.3$\times10^{-10}$\,M$_\odot$/yr   (best fit) and a 90\% confidence range 3.3-7.3$\times10^{-10}$\,M$_\odot$/yr.
The strong lines appear broadened at the 200\,km/s level. Finally, we also derive our main parameter of interest, the maximum temperature, which we use to constrain the WD mass and for which we obtain kT$_{\rm max}$=36.5\,keV, with a 90\% confidence range of 20.2--55.7\,keV.  

The presence of the complex absorber is the hallmark of magnetic CV's (Ramsay et al. 2008). In addition, the excess of counts below 0.5\,keV could indicate a soft component arising from the heated white dwarf (WD) surface. Hence, we computed the WD mass assuming that CP Pup is indeed a magnetic CV and obtained M$_1$=0.80$_{-0.23}^{+0.19}$\,M$_\odot$.  Note that for maximum temperatures as high as $\sim$36\,keV, small changes in kT$_{\rm max}$ only lead to subtle difference in the continuum in the HETG range, so we must assume that we modeled the complex absorber sufficiently accurately.
However, the best handle in the HETG data for kT$_{\rm max}$ is the emission line strength.  Plasmas with kT higher than $\sim$20\,keV adds continuum flux but little line emission in the HETG band.  Thus our constraint comes from the observed line equivalent width, and the large error range is partly due to the fact that we allowed 
the abundances of several elements to vary.  

Finally we have analyzed the X-ray data searching for possible extended emissions originating from the nova shell similarly to, for example, GK\,Per (Balman 2005). We were unable to detect any X-ray extended emission and our surface brightness profile computed on the spectra perfectly matches the simulated PSF, once the constant background level has been taken into account. 

\begin{figure*}
\centering 
\includegraphics[width=8truecm, angle=0]{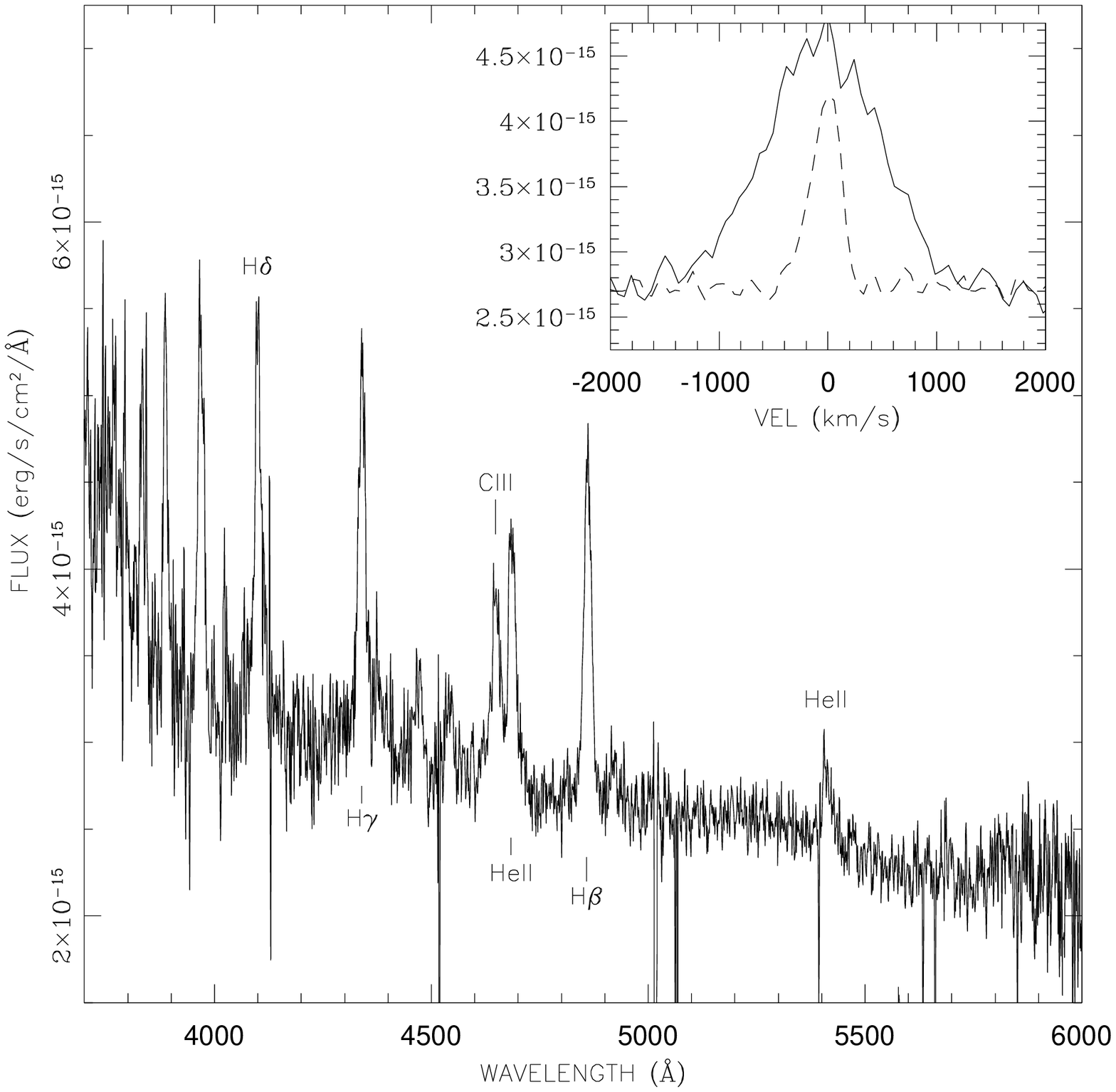}
\includegraphics[width=8truecm, angle=0]{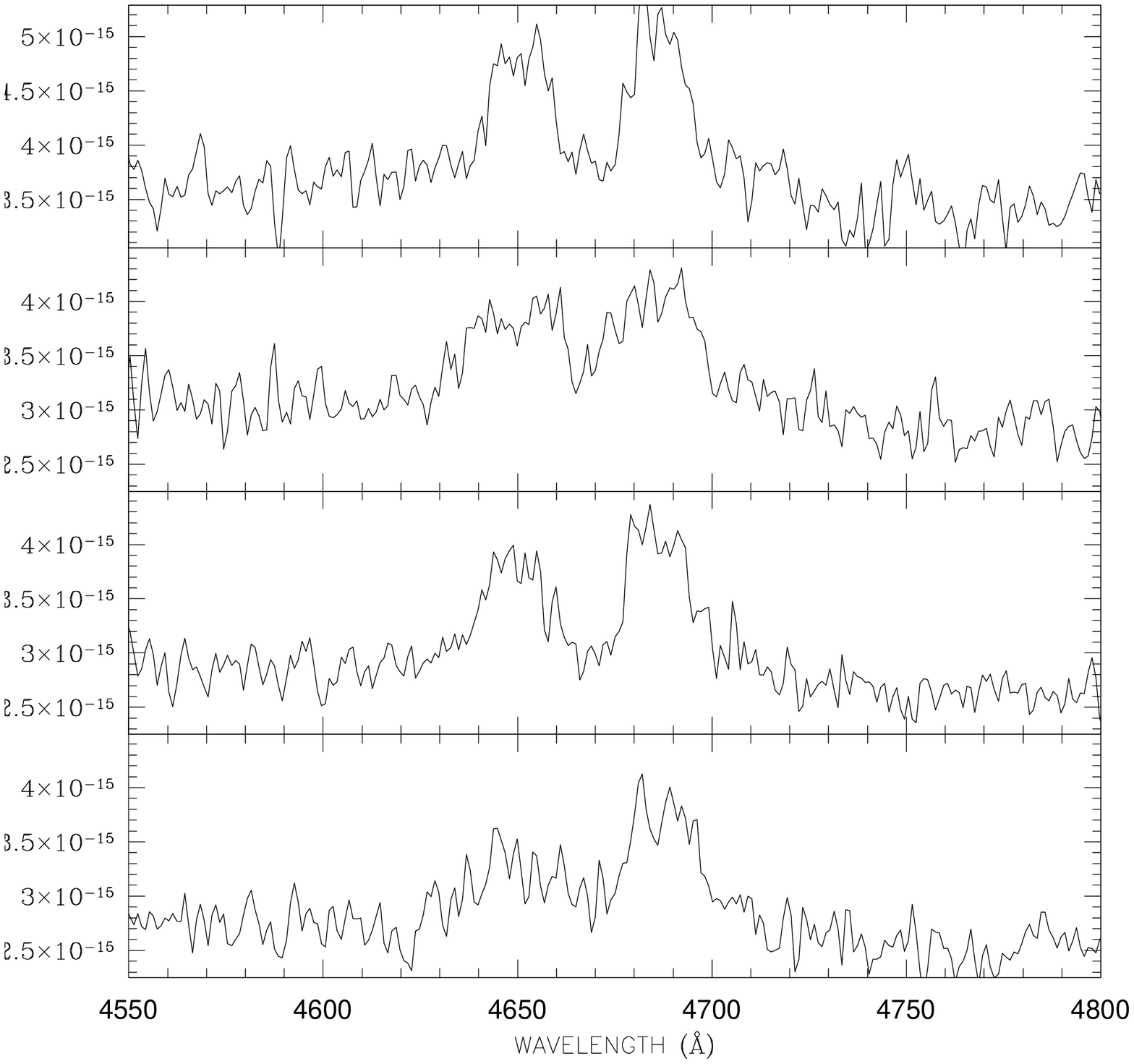}
\caption{Left panels: a representative spectrum of our time series. It has been obtained by averaging two spectra taken toward the end of the first night. Note that the spikes at 4130 4517 5018 5065 5395 and 5665 \AA\, are all bad pixels. The inset compare the H$\beta$ line profile (solid line) with that of the sky emission lines O\,{\sc i}\,$\lambda$5577 (dashed line). Right panel: example of profile variation shown by the C\,{\sc iii} and He\,{\sc ii} emission lines. The four spectra are separated by about 25-50 minutes each other. }
\label{spc}  
\end{figure*}

\begin{figure*}
\centering 
\includegraphics[width=12truecm, angle=270]{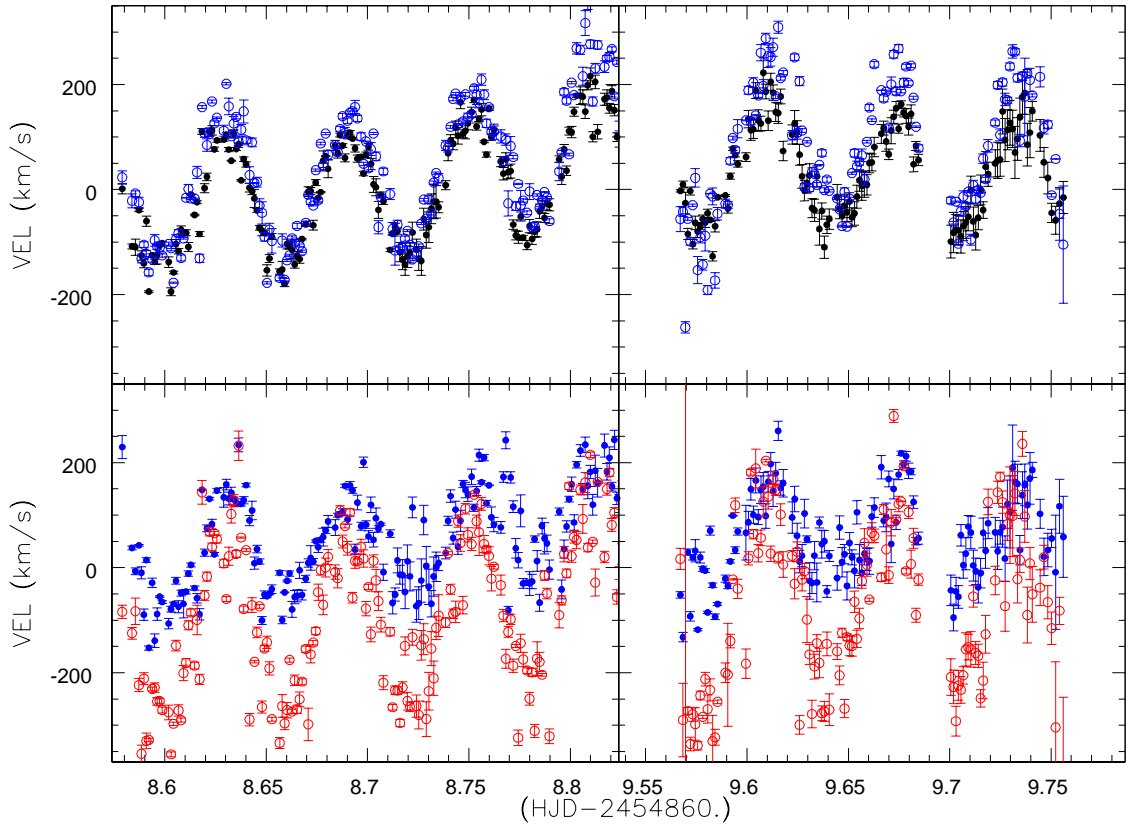}
\caption{Top panels: radial velocity curves of the emission lines H$\beta$ (black solid circles) and H$\gamma$ (blue circles) during the Feb\,6 (left) and Feb\,7 (right) observing runs. Bottom panels: radial velocity curve of the emission lines He\,{\sc ii}$\lambda$4686 (blue solid circles) and the C\,{\sc iii}+N\,{\sc iii} blend (red circles) during the same nights. The y range is the same in all the panels.   }
\label{brute_rv}  
\end{figure*} 

\section{Data analysis: the optical spectra}
We measured the wavelength of the strongest emission lines in each spectrum, i.e. the Balmer lines H$\beta$ and H$\gamma$ as well as the He\,{\sc ii}\,$\lambda$4686 and C\,{\sc iii}(1). Given the low resolution and the signal-to-noise ratio (SNR) of the spectra we could not distinguish a broad and a narrow component of the emission lines (e.g. Bianchini et al. 2012, O'Donoghue et al. 1989) and fit each emission line with a single Gaussian profile, tracking just the dominant component. We note however, that the line profile, particularly in the case of the C\,{\sc iii} and He\,{\sc ii} emissions,  occasionally deviates from Gaussian profile both because of intrinsic variations and poor SNR.  
Fig.\ref{spc} (left panel) shows the average of two spectra taken at the end of the first night; while the figure's inset, compares the profile of the H$\beta$ emission (solid line) with that of the sky line O\,{\sc i}\,$\lambda$5577 (dashed line). The right panel of Fig.\ref{spc}, instead, shows the variability displayed by the C\,{\sc iii} and the He\,{\sc ii} emission lines. 

The radial velocity (RV) curve of each line is plot in Fig.~\ref{brute_rv}. It shows that: \begin{enumerate}
 \item there is an overall drift during the first night and an offset of the average radial velocity between the two nights. 
 \item the amplitude of the $\sim$1.47\,hr period may vary from cycle to cycle.
 \item there is a large scatter of the data points with occasionally deviating points which only in part can be explained by poor SNR and/or fitting. 
\end{enumerate}


Point (i) and, in particular, the trend of the first night suggest a variability on timescale longer than our observing run. A Fourier analysis combining the whole data set of two nights gives a power spectrum that peaks at the well known period at $\sim$16.25 cycles/day ($\simeq$1.47\,hr), as well as a substantial power at low frequencies around $\sim$2.38 cycles/day (corresponding to $\sim$10.1\,hr). A formal fit to the H$\beta$ radial velocity curve gives 9.8\,hr (with the uncertainty of 1\,d, including all the aliases), which has to be regarded as a lower limit to a possible periodic variation. Although there is certain evidence of long term variability, we are unable to establish whether it is periodic or not due to the time gap between the two nights and the time coverage of our observing run.  



Points (ii) and (iii) suggest that the line forming region might be varying in location,  geometry and/or optical depth. 

\begin{figure}
\centering
\includegraphics[width=7truecm, angle=270]{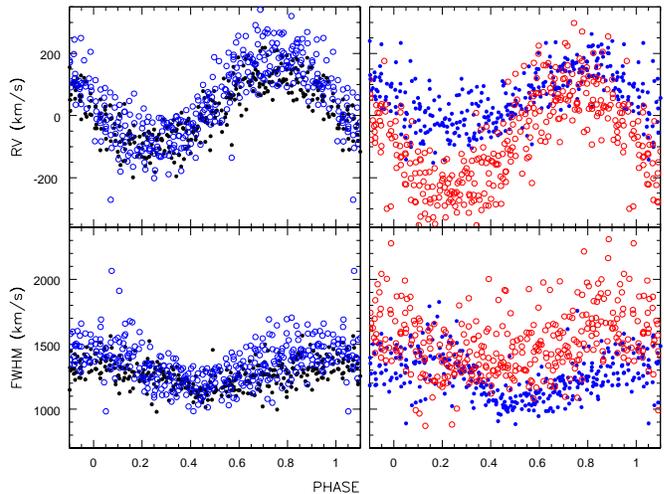}
\caption{Top panels: radial velocity curve of the emission lines H$\beta$ and H$\gamma$ (left) and He\,{\sc ii} and C\,{\sc iii} (right - same color code as in Fig.~\ref{brute_rv}). Bottom panels: modulation of the emission lines FWHM with the $\sim$1.47\,hr period. Note that while H$\beta$ and H$\gamma$ FWHM present an almost sinusoidal modulation, the He\,{\sc ii} modulation is more similar to a sawtooth profile. The C\,{\sc iii} FWHM is much less modulated and ``noisy'', possibly indicating a variable contribution/blend with the N\,{\sc iii} emission line. }
\label{spin}
\end{figure}

Phasing the spectra on the ephemeris: 
T$_0$=2452868.582093+$n\times$0.0614466\,HJD, 
which we derived from the best fit for the H$\beta$ RV curve, we observe that (see Fig.\ref{spin} top panels) different ions have different K-amplitudes, different average-RV and same red-to-blue crossing phase, suggesting a structured line forming region. 
The line dispersion velocity (FWHM) is fairly large in the range 1000-2000\,km/s and modulated across the high frequency period (see Fig.\ref{spin}, bottom panels). The modulation varies in amplitude and shape with the ions, again consistently with a structured line emitting region and possible optical depth effects.  

We remark that we did not observe modulations of the lines equivalent width (EW), nor correlation between, e.g., the He\,{\sc ii} and the H$\beta$ EWs. This possibly indicates that the line strength is larger when a more collimated area is into view and/or the presence of more than one line forming region. 
We also note that the  C\,{\sc iii} velocities and variable profile are suggestive of a significant contribution from the N\,{\sc iii}$\lambda$4640 line of $\sim$25\%.  The lack of significant modulation of its FWHM might indicate that the N\,{\sc iii} fractional contribution is phase dependent.  

\begin{figure*}
\centering 
\includegraphics[width=8truecm, angle=0]{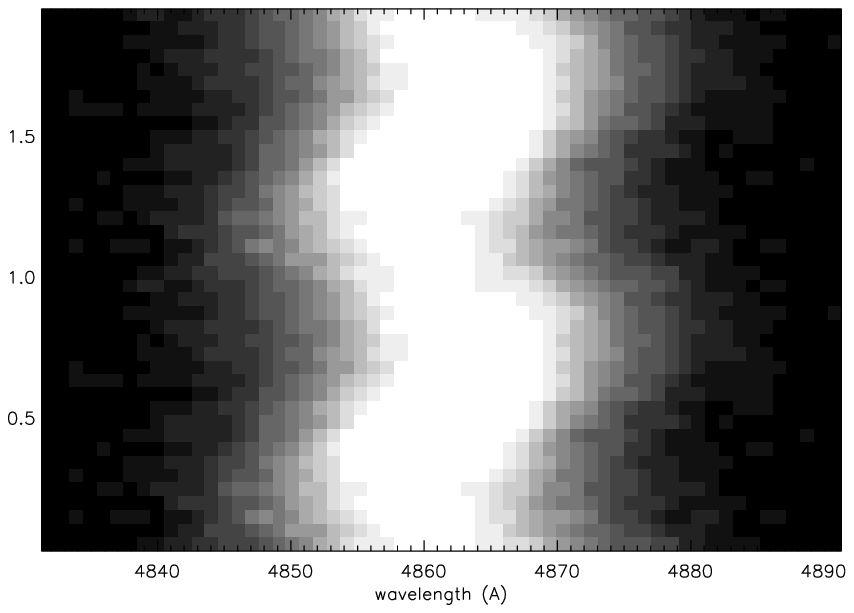}
\includegraphics[width=8truecm, angle=0]{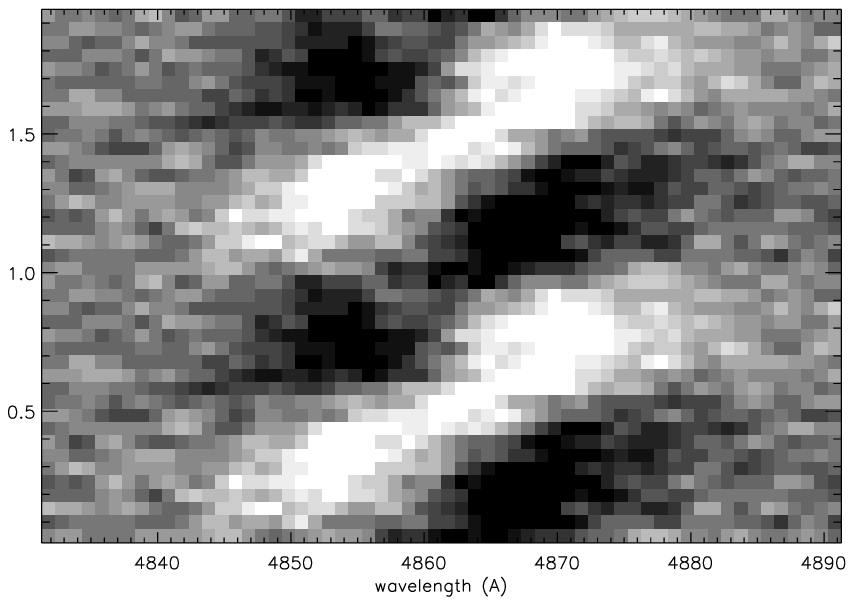}
\includegraphics[width=8truecm, angle=0]{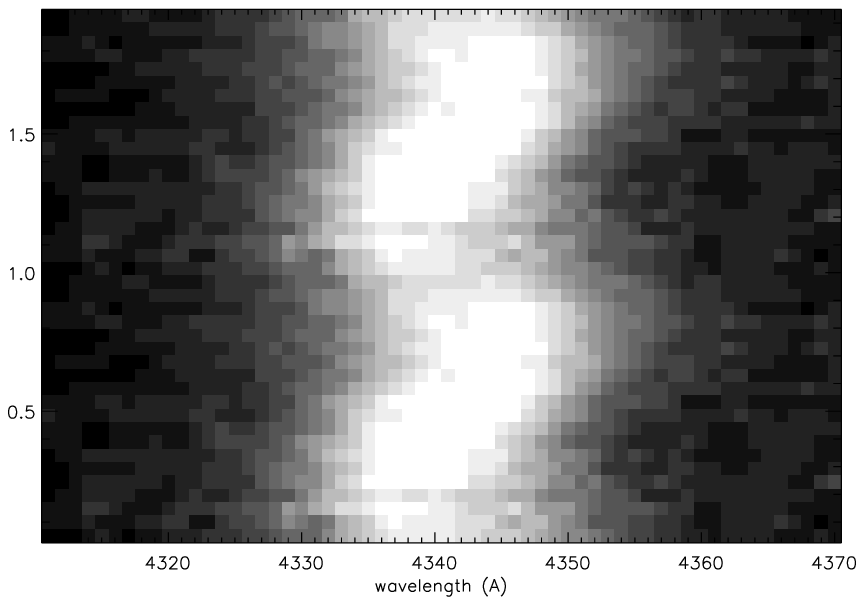}
\includegraphics[width=8truecm, angle=0]{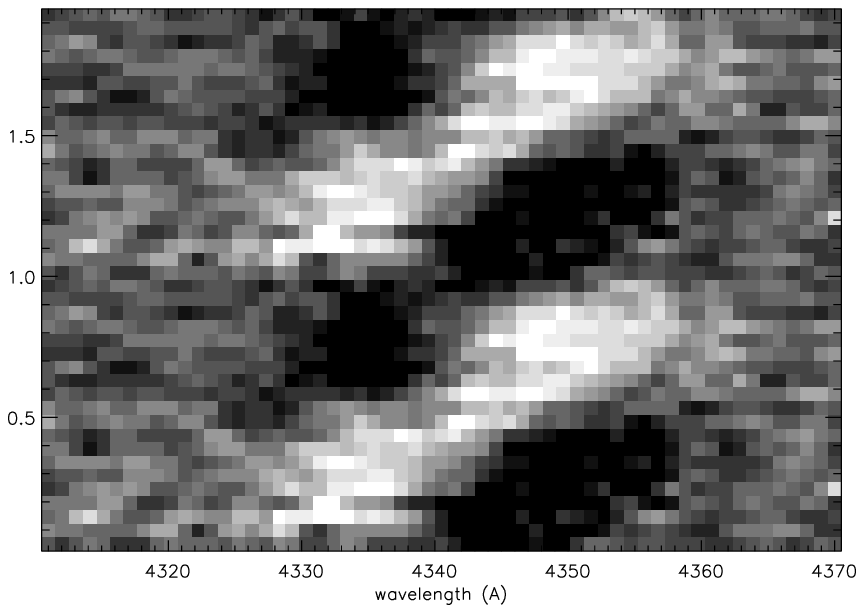}
\includegraphics[width=8truecm, angle=0]{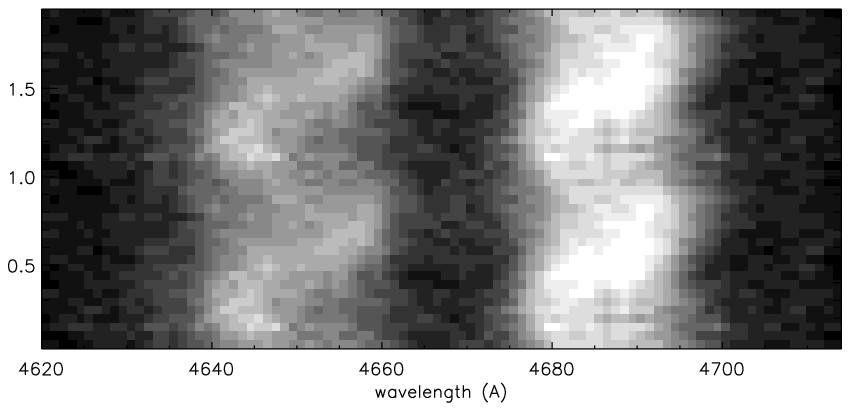}
\includegraphics[width=8truecm, angle=0]{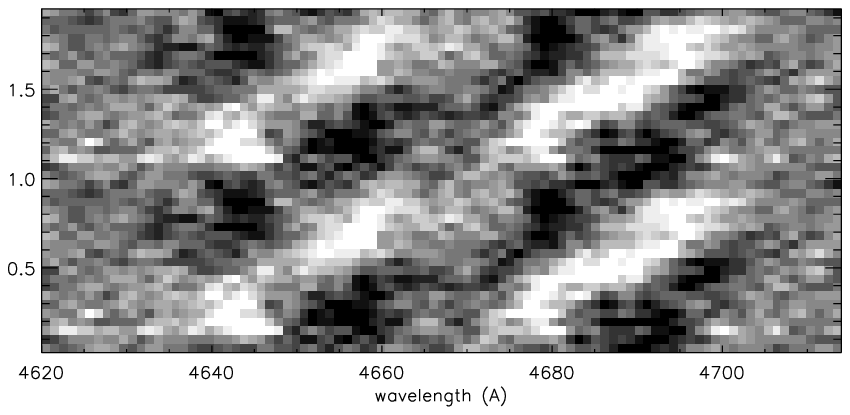}
\includegraphics[width=8truecm, angle=0]{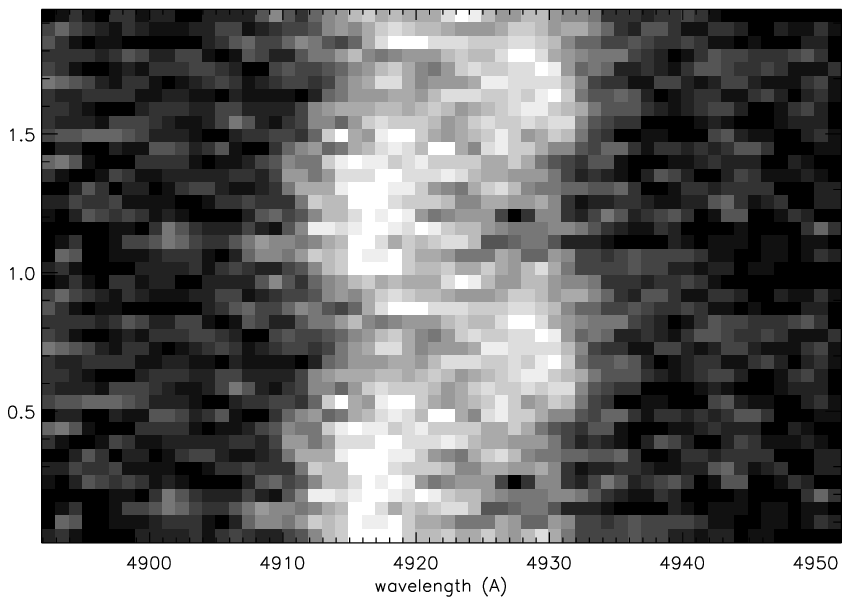}
\includegraphics[width=8truecm, angle=0]{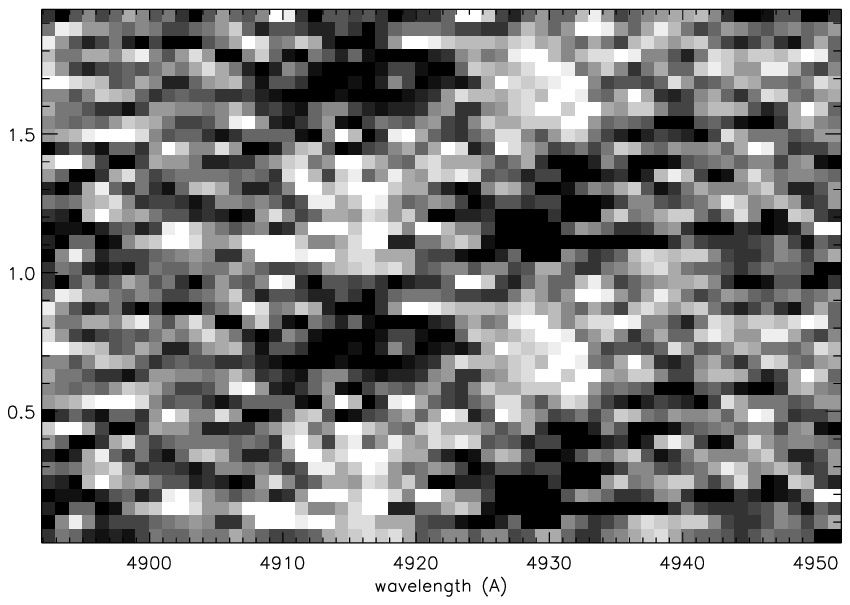}
\caption{Trailed spectra (left panels) and ``difference trailed spectra'' (right panels) of the emission lines H$\beta$, H$\gamma$, He\,{\sc ii} + C\,{\sc iii} and He\,{\sc i}, from top to bottom. Phase 0 corresponds to the time for the red-to-blue crossing in the high-frequency modulation.
}
\label{trail}  
\end{figure*}

Finally we plot in Fig.~\ref{trail}  the trailed spectra of the various emission lines folded on the 1.47\,hr period. In addition to the ``standard'' trailed spectra we produced ``difference trailed spectra'' by subtracting CP\,Pup average spectrum to each spectral bin. Difference trailed spectra  enhance the line asymmetries and their motion.
The trailed spectra reveal a ``helix'' pattern of the emission lines across the 1.47\,hr period which became suggestive, again, of a structured or a multicomponent line forming region particularly in the case of the He\,{\sc ii} line. 

\section{Discussion}
In Section\,1 we mentioned that the observations collected at all wavelength (X to NIR) for the quiescent CP\,Pup provide inconsistent results. On one side the X-ray observations always favored a magnetic nature of the nova: the X-ray flux is modulated at the high frequency period (Balman et al. 1995, Orio et al.  2009) suggesting that it corresponds to the WD spin period. In addition, the observed X-ray spectrum  is best modeled by  high temperature optically thin plasma that is commonly found in  magnetic systems. On the other hand, polarimetry observations (Cropper 1986) did not reveal polarized light in CP\,Pup disfavoring the polar interpretation, though not the IP scenario. However, Cropper's (1986) observations cannot be conclusive due to their short time coverage (18\,min). 
Time resolved optical observations (Bianchini et al. 1985, 2012, Warner 1985, Duerbeck et al. 1987, O'Donoghue et al. 1989, Barrera \& Vogt 1989, Diaz \& Steiner 1991, White et al. 1993, Patterson \& Warner 1998) detected unstable periods $<$2\,hr. In particular, spectroscopic periods in the range 0.06115-0.06148\,days, were interpreted as the binary orbital period, while photometric periods in the range 0.06138-0.0614\,days were interpreted as super-humps, as commonly observed in short-orbital period systems accreting at high rates (Patterson \& Warner 1998). In view of these periods CP\,Pup was recognized as the first CN below the period gap. 
However, the WD mass value determined from optical RV curves is uncomfortably low, in the range 0.02$\leq$M$_1<$0.6\,M$_\odot$ depending on the author and the adopted orbital inclination (White et al. 1993, O'Donoghue et al. 1989, Duerbeck et al. 1987, Barrera \& Vogt 1989, Bianchini et al. 2012). 
Only Bianchini et al. (2012) were able to force an analytical solution with M$_1\simeq$1\,M$_\odot$ by including a peculiar hot spot contribution to the line flux. However, no past, nor present observations are consistent with the configuration of accretion disk +  hot spot. 
WD masses $<$0.6\,M$_\odot$ are hardly compatible with the classical nova (CN) theory and, in particular, with the fast super-Eddington outbursts of the type displayed by CP\,Pup.   These outbursts require more massive WDs, typically$\geq$0.8-1.0\,M$_\odot$ (e.g. Starrfield et al. 2012a, 2012b, and reference therein).
Finally the NIR J-band light curve shows a double humped modulation at about the photometric/spectroscopic period, which, if interpreted as ellipsoidal variations, implies a secondary star that is not compatible with a short orbital period system below the period gap and at the distance\footnote{Calculated from the nebular parallax (Williams 1982).} of 1.6\,kpc of the nova. CP\,Pup NIR magnitudes are also incompatible with the binary evolutionary phase at the measured distance, being far too bright.

Our new Chandra and CTIO observations support the magnetic nature of CP\,Pup.
The Chandra data show evidence of a multi-temperature plasma which is absorbed by dense material partially covering the X-ray source and temperatures up to kT=36.5\,keV. High plasma temperatures and complex local absorbing medium are characteristic of mCVs (Ramsay et al. 2008). Non-mCVs do not suffer strong local absorptions except for high orbital inclination systems (e.g. the eclipsing system V893\,Sco, Mukai et al. 2009). 
Note that only the Chandra HETG has the sensitivity, band-pass and spectral resolution 
 that allow to assess the presence of a complex absorber.  
Note also that the WD mass can be derived from the maximum plasma temperature, kT$_{\rm max}=36.5$\,keV, either regarding it as shock temperature onto a magnetized WD (i.e. free-fall), or as a temperature produced by accretion from a disk onto a non-magnetized WD (i.e. Keplerian velocities). In the latter case the kT$_{\rm max}$ temperature would be reached by a more massive WD: M$_1=1.1^{+0.15}_{-0.36}$\,M$_\odot$ (as opposite to the M$_1=0.8^{+0.19}_{-0.23}$\,M$_\odot$ in the magnetic WD hypothesis). 
Both values, together with their uncertainties fall within the typical CV primary masses (Zorotovic et al. 2011) and point to a higher WD mass than previously  estimated which is more consistent with the CN theory.

Our sequence of optical spectra imposes some strong constraint on the location and geometry of the line forming region suggesting a composite  line forming region with a temperature gradient off the orbital plane. 
In addition, the observed $\gamma$-drift implies either that the RV are affected by non orbital mechanisms, or the existence of a longer period (yet to be confirmed). Note that the $\gamma$-drift is not just a one night occurrence as it appear also in the re-examination of the older data sets (Bianchini et al. 1985, 2012). It is intriguing to notice, however, that we mostly detect positive drifts. Whether this is due to our specific sampling or is intrinsic to the line emitting source cannot be concluded at this stage, nor we can say whether the phenomenon is strictly periodical. Anyway, the lines must form in the binary accretion component. They cannot, for example, form on the irradiated face of the secondary star for their FWHM are too large to be consistent with such a hypothesis and the emitting secondary cannot explain the $\gamma$-drift. In addition, the broad and narrow components identified by O'Donogue et al. (1989) and Bianchini et al. (2012) and which we do not resolve, appear to be in phase arguing, too, against secondary star irradiation. 
The accretion component can either be an accretion disk in a non-mCV or the magnetically funneled gas in an mCV. Below we discuss the two scenarios separately in the attempt to identify which one apply to CP\,Pup.

\subsubsection*{non-mCV}

The classic disk + hot-spot configuration explains the different $\gamma$ velocities from the different lines, but not their matching phases, nor the $\gamma$-drift. The hot-spot typically shows temperature gradients along the stream trajectory on the orbital plane imposing significant phase offsets among different lines (e.g. Mason et al. 2000 and reference therein), contrarily to CP\,Pup observations. It is possible to observe different $\gamma$ velocities for a same line  at different epochs. This is explained with long term variation of the geometry and/or velocity field of the line forming region, possibly associated to changes in the mass transfer rate.  
However, ``real time variations'' as those observed in CP\,Pup, if due to sudden mass transfer rate changes, should be associated to significant line intensity and profile variations as observed during flares or episodes of intermittent mass transfer (e.g. Mason et al. 2007, Rodriguez-Gil et al. 2012). We observe only small scale variations which are consistent with the period modulation and the night conditions. 
An eccentric and precessing accretion disk could explain the $\gamma$-drift and, in fact, produce a sinusoidal modulation of the $\gamma$ velocity. The precessing disk would be consistent with Patterson \& Warner (1998) super-hump proposal to explain the light curve modulations, though their instabilities (the ``anomalous super-hump'', as defined by Patterson \& Warner 1998), remain quite unexplained. We note that this model is no longer valid should future observations prove that the $\gamma$ modulation follow a  ``sawtooth'' pattern instead of a sinusoidal one. 
The exact location of the line forming region on the disk + hot-spot, however, remain unclear as there is no obvious configuration for having smaller K-amplitude in case of higher ionization potential ions. 
The observed FWHM are quite large for a low orbital inclination system and are hard to explain as are their modulation.  Our trailed spectra do not match the pattern displayed by disk + hot-spot or hot-spot emission. 

\subsubsection*{mCV}

First of all,  the presence of strong He\,{\sc ii} emission lines is a strong indicator of the magnetic nature of the CV (van Paradijs \& Verbunt 1984). Second, in a mCV the bulk of the line emitting region is located in the magnetically confined accretion flow toward the WD magnetic poles (Schwope et al. 2000). This configuration naturally imposes an off-orbital-plane/vertical structure characterized by temperature and velocity gradients and non-Keplerian velocities. The various emission lines are displaced along the curtain and therefore their RV curves have matching phases but different $\gamma$ velocities. The higher ionization potential energy transitions form closer to the magnetic pole and the K-amplitude distribution results from a projection effect. Similarly the line FWHMs and their modulation are imposed by the velocity field along the magnetic field lines and its projection on the line of sight and consistent with those expected in IP (Warner 1995 and reference therein, Ferrario et al. 1993). The helix pattern shown by the trailed spectra suggest multiple components of the type observed in mCV and IP in particular (see e.g. Hellier 1999). Finally, the $\gamma$ velocity drift suggests two different possible scenarios:\\ 
1) the modulation is sinusoidal and induced by the orbital motion of the WD around the binary center of mass. I.e. the $\sim$1.47\,hr period is the WD spin and the true orbital period is several hours long (similarly to RXJ154814.5-452845, see, e.g. de\,Martino et al. 2006) and has yet to be determined; while the photometric period is a beat. \\
2) the $\gamma$ modulation is asymmetric and sawtooth-like  and the line emitting region is affected by cyclical disturbances.  \\
We note at this point that neither the WD spin, nor the binary orbital period are expected to be unstable. However, they can appear as such if their probe is unstable. 
The accretion curtain (or threading region) is fixed in the binary frame if the system is synchronous; it is not otherwise. The curtain will have constant geometry and emissivity if it is fed by a symmetric accretion disk/ring, but will vary in case of asymmetries within the disk (either because the disk is missing or because the accretion stream is deeply penetrating into it). Hence the observer will view a line emitting region which is modulated with the WD spin and its beat. 
We note to this regard that different accretion geometries are possible in IPs, depending on the magnetic field strength and the ratio between the spin and the orbital period (Warner 1995 and reference therein, Norton et al. 2004). 
We also note that the magnetic system BY\,Cam shares a number of similarities with CP\,Pup. It has a record of multiple and non repeatable periods with no optical nor polarimetry observations matching either the spin or the orbital period and depending on the sampling (e.g. Silber et al. 1997 and reference therein, Mouchet et al. 1997). Both light and RV curves show a phase shift and an abrupt reset to 0 which has been explained with the change in the magnetic geometry seen by the accretion stream as the WD rotates relative to the secondary and the switching of accreting magnetic pole\footnote{Mouchet et al. (1997) invoke a precessing WD, but in facts this is unnecessary. At the same time a similar effect could be obtained without invoking the change of accreting pole but the reconnection of the magnetic field line from the opposite side once the WD magnetic pole has swept over 180$^{\rm  o}$ with respect to the L1 point.}. It is not difficult to imagine that similar configuration could cause a drift in $\gamma$ velocity beside the phase. 

Independently on the value of the orbital period, the double humped NIR modulation might be explained with irradiation (of the secondary star or the accretion component, e.g. Penning 1985) by two accreting magnetic poles. Note that CP\,Pup is a low orbital inclination binary, hence, we might very well miss the direct view of the lower magnetic pole (or see its periodic eclipse while the upper pole is always visible), but still see its effects on the secondary (/accretion component). Strictly speaking the double humped modulation in this case should result in uneven maxima, however this is hard to say from Szkody \& Feinswog (1988) plot due to the paucity of data points and the lack of error bars. Most important those data cannot be conclusive on irradiation (nor on ellipsoidal variations), for their low amplitude modulation might be not significant and their phasing with respect to optical photometry is unknown. 
Note also that lack of detection of secondary star features does not argue against CP\,Pup as a possible long orbital period system. On one hand, the system spectroscopic observations have always been limited to the He\,{\sc ii} -- H$\beta$ wavelength range, which is devoid of secondary star absorptions, and to low spectral resolutions, which smear out any possible weak absorption feature. Our CTIO spectra do not have the sufficient spectral resolution and SNR, too. On the other hand, the chance for detecting the secondary depends also on the temperature and size of the various emission sources into play and it might well be that the whole accretion component completely outshine a relatively cold secondary star. 

In summary, while we cannot be conclusive on the strictly-periodic vs semi-periodic nature of the observed $\gamma$-drift and its modulation, comparing the existing observational characteristics of CP\,Pup within the mCV and non-mCV scenarios, the nova IP nature best fit them all by far. This is shown in Table~\ref{scenarios} where we list CP\,Pup observational facts and how each scenario succeed or fail in explaining those consistently.

For completeness we should also consider that the  possible existence of a longer period could imply that CP\,Pup is in fact a triple system, i.e. a CV below the period gap ``orbiting around'' a third body. In this case the hypothesis of an eccentric orbit around the third body would be needed if the asymmetry of the long term RV modulation with slow/long-lasting velocity increase and short/rapid velocity decreases, is confirmed.  
This scenario however is somewhat cumbersome and ad hoc and would open questions about its probability and stability. It would probably requires formation within a globular cluster with subsequent ejection, an evolutionary scenario which, beside being very unlikely, needs detailed simulation (see, e.g.  Anosova 1996, Brasser 2002) that are not possible with the current set of information in hand. Hence we do not pursue this scenario any longer and note that 1) the only CV that has been suggested to be part of a triple system is FS\,Aur with a period around the third body of few hundreds of days (Chavez et al. 2012); 2) placing a non-mCV in a triple system we will face the same contradictions and problem we are discussing for the isolated non-mCV. 


\begin{table*}
\centering
\scriptsize
\caption{\label{scenarios}  Comparison of the possible scenarios applicable to CP\,Pup: the non magnetic CV, the IP (or asynchronous polar), both as long or short orbital period system. The first column lists the observed characteristics and for each model we report whether it fits the observation (Y) or not (N) or whether it fits them just partially ($\sim$). We also describe how the model explains/fits the observation or why it does not, as appropriate. } 
\begin{tabular}{p{3.5cm}p{4.cm}p{4.cm}p{4cm}} 
\hline
{\sc observation} & {\sc non-mCV} & {\sc mCV/IP} & {\sc mCV/IP} \\
                  & short period & long orbital period & short orbital period  \\
\hline 
strong He\,{\sc ii}$\lambda$4686 & [N]: quiescent (low $\dot{m}$) non-mCV typically do not show He\,{\sc ii} emission & [Y]: strong He\,{\sc ii} emission line is a characteristic of mCVs & [Y]: strong He\,{\sc ii} emission line is a characteristic of mCVs \\
\hline
long term trend  & [Y]: the drift in $\gamma$ is induced by a precessing eccentric disk. 
& [Y]: the observed changes in the $\gamma$ velocity corresponds to the true (still undetermined) orbital period of CP\,Pup; the $\sim$1.4\,hr period is the WD spin. & [Y]: the $\gamma$ drift is induced by the varying geometry of the magnetic field lines as seen by the accretion stream due to the WD rotation.  \\
\hline
NIR light curve &  [N]: ellipsoidal variations due to the elongated secondary star. Ellipsoidal variations are inconsistent with both the distance and the evolutionary phase of the binary.  & [Y]: the spinning WD magnetic poles outshine twice per WD rotation the secondary star (/accretion component) which reflects and reprocesses the X-ray similarly to DQ\,Her. & [Y]: as for the long orbital period mCV scenario. \\
\hline
Difference between spectroscopic and photometric period (by a few \%) & [Y]: the photometric period is the super-hump period, while the spectroscopic period is the orbital one. & [Y]: beat between spin and orbital period. & [Y]: phase jitter and unstable periods as observed in BY\,Cam.  \\
\hline
dynamical WD mass & [N]: the primary masses derived from dynamical solutions assuming that the optical emission lines trace the binary orbital period are always too low to be consistent with stellar evolutionary theory and/or the CN outburst theory. & [Y]: reasonable masses could be found with the longer orbital period. & [Y]: the emission line region simply does not trace the orbital motion of the binary and cannot be used for the computation of dynamical masses. \\
\hline
X-ray spectrum & [N]: the observed spectral energy distribution and high temperature match typical values of mCVs. & [Y]: the spectral characteristic are those of an mCV & [Y]: as for the long orbital period mCV scenario. \\ \hline
presence of complex absorber in the high energy X-ray & [N]: this is not typical of non-mCV and, in particular, low orbital inclination non-mCVs & [Y]: this is an hallmark of mCVs & [Y]: this is an hallmark of mCVs \\
\hline
X-ray light curve & [N]: in non-mCV disk systems the X-ray emission is from the boundary layer which is supposedly symmetric producing no modulations. & [Y]: the X-ray modulation is consistent with mCVs. The modulation result from the occultation of one magnetic pole. & [Y]: as for the long orbital period scenario. \\
\hline 
modulation of the high frequency period (both photometric and spectroscopic) & [$\sim$]: anomalous super-humps & [Y]: phase jitter and period instability induced by asynchronism between the WD spin and the orbital period, similarly to BY\,Cam. 
& [Y]: as for the long orbital period scenario. 
\\
\hline
modulated amplitude (spectroscopic and photometric) & [$\sim$]: anomalous super-humps & [Y]: variable magnetic field geometry as viewed from the secondary star and the accretion disk, due to the WD rotation and chaotic accretion.  & [Y]: as for the long orbital period scenario.
\\
\hline
undetected circular polarization & [Y]: non-mCV do not show polarized light & [Y]: IP polarization is typically masked by the disk/accretion component emission$^\dagger$ & [Y]: as for the long orbital period scenario. \\
\hline
undetected secondary star features & [Y]: the secondary is a cool late type main sequence star as expected for short orbital period systems. Note that this is ``self-contradictory'' with the explanation provided for the non-mCV scenario on the NIR light curve point. & [$\sim$]: the secondary star should be relatively large and possibly competing in size (but not in temperature) with the accretion component. It should be searched for at NIR wavelengths. & [Y]: as for the non-mCV scenario.  \\
\hline
\end{tabular}
\scriptsize
\\
$^\dagger$ Note that the non-detection of polarized light favor the IP scenario more than the asynchronous polar one. \\
$^\ddag$ Note that a long orbital period does not automatically implies that the secondary is visible in the R, I or NIR bands: high mass transfer rate and large/luminous accretion disk can mask the secondary star absorptions. 
\end{table*}

\section{Summary and conclusions}

In this paper we presented and analyzed {\sl Chandra}/HETG X-ray
spectra and high cadence, time resolved optical spectra obtained
 with the Blanco telescope at CTIO. 
The two data sets together produce interesting results,  
demonstrating that CP\,Pup is very likely to be an IP system.
 The X-ray spectral characteristics are consistent
 with those of known IP's. 
Assuming that CP Pup hosts
 a magnetic WD we fitted the X-ray spectrum with  
a cooling-flow model and a complex absorber,
 and from the best fit we derived
 a mass of 0.8\,M$_\odot$ for the WD, which is in reasonable agreement with the typical WD masses
 in CV and with the prediction from the CN outburst theory.
The IP hypothesis thus solves the problem of the non-realistic low mass derived
 assuming that accretion occurs
 only through a disk.

Although in Orio et al. (2009) some of us derived an upper limit of
 1.6 $\times$ 10$^{-10}$ M$_\odot$ year$^{-1}$ for the 
mass accretion rate $\dot m$ onto the CP Pup WD, the {\sl Chandra} HETG spectrum indicates
 an $\dot m$ 
 {\it upper limit} of 3.3 $\times$ 10$^{-10}$ M$_\odot$ year$^{-1}$ within a 90\% confidence
 level, which may be too high to develop a very luminous TNR (Yaron et al. 2005) and
 leaves open the possibility of a variable $\dot m$ before and after the outburst.

 Like the X-ray data,
also the sequence of optical spectra,  the emission line characteristics and their radial velocities 
are much better explained with the IP scenario rather than with a non-mCV.
 We also observed a drift of the system $\gamma$ velocity,
 which is either suggestive the presence of a cyclical non Keplerian mechanism or of a longer period.
 We speculate that  this ``hidden'' period may be the true orbital
 one, rather than the spectroscopic period that may  
 be associated with the WD rotation. 
 While with the current set of data in hand we cannot be conclusive about the origin of the
 observed $\gamma$ velocity drift, we conclude that we found very strong evidence
that CP Pup is an IP.   

New optical observations are needed in order to solve the pending issues. 
High-cadence time resolved spectroscopy for several consecutive nights 
would allow to
 determine the period and the morphology of the $\gamma$ velocity modulation.
The number of nights we requested at the Blanco was 3, but we lost one due to weather,
 and it turns out that a run consisting of more nights ($\geq$4) is necessary. A simultaneous runs of broad band, multicolor photometry done  
 with another telescope should allow to assess  correlations
 between light and RV curves,  to observe
whether the continuum and the line emission originate in the same
 region, and determine how many
different sources of optical  exist.
 High resolution spectra are
desirable to properly identify and trace the various emitting components.
A long monitoring photometric program should be undertaken to best
 identify the beat periods into play. While we are aware that for this ideal dataset
  many nights of observations with telescopes of different sizes and
 characteristics would be necessary, we do think that CP Pup is such an emblematic object
 for the nova theory that obtaining these results would be very rewarding for a full
 understanding of novae.


\section*{Acknowledgments}
AB thanks the Space Telescope Science Institute (STScI) for the kind hospitality in June 2012, when this work started. EM is grateful to Steven Shore for the always stimulating conversations. The authors are thankful to the anonymous referee for having raised important points.


\begin{thebibliography}{99}

\bibitem[\protect\citeauthoryear{}{1996}]{} Anosova, J., 1996,  Ap\&SS, 238, 223

\bibitem[\protect\citeauthoryear{}{1995}]{}Balman, S., Orio, M., \&  Oegelman, H., 1995, ApJ, 449, L47

\bibitem[\protect\citeauthoryear{}{1989}]{}Barrera, L.H., \& Vogt, N., 1989, RMxA.19, 99

\bibitem[\protect\citeauthoryear{}{1985}]{}Bianchini, A., Friedjung, M., \& Sabbadin, F., 1985, Inf. Bull. Variable Stars, 2650 

\bibitem[\protect\citeauthoryear{}{2012}]{}Bianchini, A.; Saygac, T.; Orio, M.; Della Valle, M.; Williams, R., 2012, A\&A, 539, 94

\bibitem[\protect\citeauthoryear{}{2002}]{} Brasser, R., 2002, MNRAS, 332, 723

\bibitem[\protect\citeauthoryear{}{1979}]{}Cash., W.; 1979, ApJ, 228, 939

\bibitem[\protect\citeauthoryear{}{2012}]{} Chavez, C.E., Tovmassian, G., Aguilar, L. A., Zharikov, S., Henden, A.A., 2012, A\&A, 538, 122

\bibitem[\protect\citeauthoryear{}{1986}]{}Cropper, M., 1986, MNRAS, 222, 225

\bibitem[\protect\citeauthoryear{}{2006}]{} de Martino, D., Bonnet-Bidaud, J.M., Mouchet, M., Gaensicke, B.T., Halberl, F., Motch, C., 2006, A\&A, 449, 1151

\bibitem[\protect\citeauthoryear{}{1991}]{}Diaz, M. P., \& Steiner, J. E., 1991, PASP, 103, 964

\bibitem[\protect\citeauthoryear{}{1987}]{}Duerbeck, H. W., Seitter, W. C., \& Duemmler, R., 1987,  MNRAS, 229, 653 


\bibitem[\protect\citeauthoryear{}{1993}]{}Ferrario, L., Wickramasinghe, D. T., \& King, A. R., 1993, MNRAS, 260, 149

\bibitem[\protect\citeauthoryear{}{1999}]{} Hellier, C., 1999, ApJ, 519, 324

\bibitem[\protect\citeauthoryear{}{2005}]{} Kalberla, P.M.W., Burton, W.B., Hartmann, Dap, Arnal, E.M., Bajaja, E., Morras, R., Pöppel, W.G.L., 2005, A\&A, 440, 775 

\bibitem[\protect\citeauthoryear{}{2010}]{} Kalberla, P.M.W., McClure-Griffiths, N.M., Pisano, D. J. Calabretta, M. R., Ford, H. Alyson, Lockman, Felix J., Staveley-Smith, L., Kerp, J., Winkel, B., Murphy, T., Newton-McGee, K., 2010, A\&A, 512, A14

\bibitem[\protect\citeauthoryear{}{2000}]{} Mason, E., Skidmore, W., Howell, S.B., Ciardi, D.R., Littlefair, S., Dhillon, V.S., 2000, MNRAS, 318, 440

\bibitem[\protect\citeauthoryear{}{2007}]{} Mason, E., Wickramasinghe, D., Howell, S.B., Szkody, P., 2007, A\&A, 467, 277


\bibitem[\protect\citeauthoryear{}{2000}]{} Mouchet, M., Bonnet-Bidaud, J.M., Somov, N.N., Somova, T.A., 1997, A\&A, 324, 109

\bibitem[\protect\citeauthoryear{}{2003}]{}Mukai, K.; Kinkhabwala, A.; Peterson, J. R.; Kahn, S. M.; Paerels, F., 2003, ApJ, 586L, 77

\bibitem[\protect\citeauthoryear{}{2009}]{}Mukai, K.; Zietsman, E.; Still, M., 2009, ApJ, 707, 652

\bibitem[\protect\citeauthoryear{}{2004}]{} Norton, A.J., Haswell, C.A., Wynn, G.A., 2004, ApJ, 614, 349

\bibitem[\protect\citeauthoryear{}{1989}]{}O'Donoghue, D., Warner, B., Wargau, W., Grauer, A.D.,1989,  MNRAS, 240, 41 

\bibitem[\protect\citeauthoryear{}{2009}]{} Orio, M., Mukai, K., Bianchini, A., de Martino, D., Howell, S.B., 2009, ApJ, 690, 1753

\bibitem[\protect\citeauthoryear{}{1985}]{}Penning, W. R., 1985, ApJ, 289, 300

\bibitem[\protect\citeauthoryear{}{1998}]{}Patterson, J., \& Warner, B., 1998, PASP, 110, 1026

\bibitem[\protect\citeauthoryear{}{1984}]{}Paradijs, D., \& Kovetz, A., 1995, ApJ, 445, 789

\bibitem[\protect\citeauthoryear{}{2008}]{}Ramsay, G.; Wheatley, P.J.; Norton, A.J.; Hakala, P.; Baskill, D., 2008, MNRAS, 387, 1157

\bibitem[\protect\citeauthoryear{}{2012}]{} Rodríguez-Gil, P., Schmidtobreick, L., Long, K.S., Gaensicke, B.T., Torres, M.A.P., Rubio-Díez, M.M., Santander-García, M., 2012, MNRAS, 422, 2332

\bibitem[\protect\citeauthoryear{}{2010}]{}Schaefer, B. E.,  \& Collazzi, A. C., 2010, AJ, 139, 1831 

\bibitem[\protect\citeauthoryear{}{2000}]{} Schwope, A.D., Catalan, M.S., Beuermann, K., Matzner, A., Smith, R.C., Steeghs, D., 2000, MNRAS, 313, 533

\bibitem[\protect\citeauthoryear{}{1997}]{} Silber, A.D., Szkody, P., Hoard, D.W., Hammergren, M., Morgan, J., Fierce, E., Olsen, K., Mason, P.A., Rolleston, R., Routsalainen, R., Pavlenko, E., Shakhovskoy, N.M., Shugarov, S., Andronov, I.L., Kolesnikov, S.V., Naylor, T., Schmidt, E., 1997, MNRAS, 290, 25

\bibitem[\protect\citeauthoryear{}{1988}]{}Szkody, P., \& Feinswog, L., 1988, ApJ, 334, 422

\bibitem[\protect\citeauthoryear{}{2012}]{}Starrfield, S., Iliadis, C., Timmes, F.X., Hix, W.R., Arnett, W.D., Meakin, C., Sparks, W.M., 2012a, BASI, 40, 419

\bibitem[\protect\citeauthoryear{}{2012}]{}Starrfield, S., Timmes, F.X., Hix, W.R., Iliadis, C., Arnett, W.D., Meakin, C., Sparks, W.M., 2012b, IAUS, 281, 166


\bibitem[\protect\citeauthoryear{}{1984}]{} van Paradijs, J., Verbunt, F., 1984, AIPC, 115, 49

\bibitem[\protect\citeauthoryear{}{1985}]{}Warner, B., 1985, MNRAS, 217, 1P

\bibitem[\protect\citeauthoryear{}{1995}]{}Warner, B., 1995,  Cataclysmic Variable Stars 
                 (Cambridge:  Cambridge Univ. Press)

\bibitem[\protect\citeauthoryear{}{1993}]{}White, J. C., Honeycutt, R.K., \& Horne, K., 1993, ApJ, 412, 278 

\bibitem[\protect\citeauthoryear{}{1982}]{}Williams, R. E., 1982, ApJ, 261, 170

\bibitem[\protect\citeauthoryear{}{205}]{}Yaron, O., Prialnik, D., Shara, M.M., Kovetz, A., 2005, ApJ, 623, 398

\bibitem[\protect\citeauthoryear{}{2011}]{}Zorotovic, M.; Schreiber, M. R.; Gänsicke, B. T., 2011, A\&A, 536, 42

\end{thebibliography}
\end{document}